\documentclass[journal]{IEEEtran}
\usepackage{amsmath}
\usepackage{graphicx}
\usepackage{amssymb}
\usepackage{mathrsfs}
\usepackage{mathtools}
\usepackage{makecell,multirow}

\usepackage[ruled,vlined]{algorithm2e}

\usepackage{bm}
\usepackage{stfloats}

\usepackage{subfigure}

\usepackage{color}
\usepackage{flushend} 

\newcommand{\MyAlterDel}[1]{}
\usepackage{cite}

\begin{document}
\bstctlcite{Ref-MEC-TWC:BSTcontrol}
\makeatletter
\def\bstctlcite{\@ifnextchar[{\@bstctlcite}{\@bstctl
cite[@auxout]}}
\def\bstctlcite[#1]#2{\@bsphack
\@for\@citeb:=#2\do{%
\edef\@citeb{\expandafter\@firstofone\@citeb}%
\if@filesw\immediate\write\csname #1\endcsname{\s
tring\citation{\@citeb}}\fi}%
\@esphack}
\makeatother

%
\title{Online Optimization of Wireless Powered Mobile-Edge Computing for Heterogeneous Industrial Internet of Things}

%

\author{\IEEEauthorblockN{Hao Wu$^{1}$, Xinchen Lyu$^{2}$, and Hui Tian$^{1}$}\\
\IEEEauthorblockA{$^{1}$State Key Laboratory of Networking and Switching Technology, Beijing University of Posts and Telecommunications, Beijing, China, Email: wh9405@bupt.edu.cn and tianhui@bupt.edu.cn.\\
$^{2}$National Engineering Laboratory for Mobile Network Technologies, Beijing University of Posts and Telecommunications, Beijing, China, Email: lvxinchen@bupt.edu.cn.}\vspace{-2em}
}

\maketitle
\begin{abstract}
A spurt of progress in wireless power transfer~(WPT) and mobile edge computing~(MEC) provides a promising approach for Industrial Internet of Things~(IIoT) to enhance the quality and productivity of manufacturing.
Scheduling in such a scenario is challenging due to congested wireless channels, time-dependent energy constraints, complicated device heterogeneity, and prohibitive signaling overheads.
In this paper, we first propose an online algorithm, called energy-aware resource scheduling (ERS), to maximize the system utility comprising throughput and fairness, with consideration on both system sustainability and stability.
Based on Lyapunov optimization and convex optimization techniques, the proposed algorithm achieves asymptotic optimality for heterogeneous IIoT systems without prior knowledge of network state information (NSI).
Subsequently, we extend the ERS algorithm to a more realistic scenario where the overhead and delay of NSI feedbacks are non-negligible.
The optimal scheduling decisions of the scenario are provided, and the optimality loss on system utility under outdated NSI is analyzed.
Simulations verify our theoretical claims and demonstrate the gains of our proposed ERS algorithm over alternative benchmark schemes.
\end{abstract}

\begin{IEEEkeywords}
Wireless power transfer, mobile edge computing, heterogeneous Industrial Internet of Things, outdated information, Lyapunov optimization.
\end{IEEEkeywords}

%
\IEEEpeerreviewmaketitle

\section{Introduction}
\IEEEPARstart{I}{ndustrial} Internet of Things (IIoT) is regarded as a revolutionary approach to optimize industrial production processes and improve economic benefits \cite{bibli:IIoT3,bibli:IIoT2,bibli:IIoT}.
Ubiquitous wireless devices (WDs) in IIoT work as the sources to collect an enormous amount of data from the ambient environment.
Analysis and extraction of those data can add value to the production cycle, thus improving the efficiency and accuracy of manufacturing \cite{bibli:ambient}.
Generally, the data will be transmitted to and processed at a nearby radio access point (AP) to facilitate network model training and intelligent decisions, e.g., machine learning for feature extraction\cite{bibli:AI} and fuzzy Q-learning for load balancing \cite{bibli:BD}.
To this end, endowing APs with powerful mobile edge computing (MEC) functionalities will be a major form of IIoT scenarios \cite{bibli:IIC,bibli:edge,bibli:edge2}.

A typical use case enabled by MEC in IIoT is warehouse environmental monitoring.
In such an application, different types of WDs are deployed around the storehouse, responsible for collecting and transmitting data from surroundings to APs in order to maintain a favorable storage environment.
Therefore, the system would prefer high throughput with reliability and practicality but pays less attention to latency.
Different from conventional MEC networks which focus on radio/computing resources optimization of homogeneous devices with real-time network state information (NSI), scheduling in the considered IIoT scenario faces several technical challenges:

1) \emph{Sustainability}:~The IIoT devices are increasingly empowered with wireless power transfer~(WPT) capability, so as to harvest energy from wireless signals, recharge their limited battery, and prolong the lifetimes of the devices.
This is critical to guarantee the sustainability with WPT capability since frequent replacement of batteries for a massive number of devices would incur prohibitive operational/maintenance cost in IIoT. However, the transmission of wireless power consumes spectrum resources and needs to be jointly optimized with the decisions of data offloading and processing in the AP.

2) \emph{Heterogeneity}:~The IIoT devices (such as industrial sensors, actuators, and controllers) have heterogeneous computing/storage/battery capabilities. Such inhomogeneity would necessitate specific and meticulous algorithms for different types of devices.

3) \emph{Scalability}:~Given the stochastic and time-varying features of the IIoT environment, real-time NSI is vital for the effectiveness of resource allocation.
However, due to limited signaling resources, it may not be practical for the AP to acquire real-time NSI from a wide range of IIoT devices. The aforementioned joint design under heterogeneous IIoT environment is required to be scaled to operate in the presence of outdated NSI.

\subsection{Related Work}
These challenges have yet to be addressed in the literature, especially in heterogeneous IIoT.
Although there are extensive works about resource scheduling in MEC~\cite{bibli:arrival,bibli:MEC1,bibli:MEC2}, the joint optimization for bandwidth, computing, and energy intake/output is of insufficient study.
Offline algorithm for the joint optimization would require complete non-causal information of networks and may suffer from the \emph{curse of dimensionality} when the system is in large scale.
To address these issues, a perturbed Lyapunov technique was employed to decouple the spatial- and time-dependency of multiple resources in \cite{bibli:channel} and \cite{bibli:delay}.
However, these analyses may not be suitable for IIoT, since their energy harvesting processes were modeled as stochastic energy packet arrivals.
The highly variable of energy supplies was unfavorable for system control and manual management.

Recent years have witnessed the possibility of integrating WPT with MEC.
WPT, especially in the form of wireless powered communication network (WPCN)~\cite{bibli:protocol2,bibli:noise1,bibli:noise2,bibli:globecom,bibli:CSI,bibli:type}, has been envisioned as an promising paradigm to provide adaptive and sustainable energy supply for battery-powered devices.
In~\cite{bibli:delay3}, a computation rate maximization problem was formulated in wireless powered MEC with binary offloading, and solved by developing two algorithms based on coordinate descent and alternating direction method of multipliers.
Reference~\cite{bibli:WMEC1} explored the benefits of user cooperation in minimizing WPT-MEC system energy by solving a min-max problem through a two-phase method.
For the scenario with multi-antenna AP, a semiclosed solution for AP's energy consumption minimization was derived under a computation latency constraint~\cite{bibli:multi}.
However, these algorithms implicitly assumed that the AP has sufficient computing capability.
In contrast, the optimal design for resource-limited AP, which is widespread in IIoT due to stringent production cost consideration, and its sustainability analysis are lack of study.

There have been separate studies over the optimization of WPT/MEC for the devices with limited~\cite{bibli:protocol2,bibli:protocol,bibli:multi} (referred to as Type-I hereafter) or sufficient~\cite{bibli:noise1,bibli:globecom,bibli:WMEC1,bibli:noise2} (referred to as Type-II in the sequel) battery sizes.
Type-I devices, with high self-discharge rate, consume all the harvested energy for data offloading in each time slot.
Therefore, its resource scheduling usually works in a myopic manner.
On the contrary, so long as the battery capacity allows, Type-II devices can store the unused energy harvested in the current slot for future use.
Some papers noticed the heterogeneity and proposed separate scheduling approaches for the two types of WDs~\cite{bibli:CSI,bibli:type}.
However, none of the existing works can be extended to the considered heterogeneous IIoT system, since the joint optimization for Type-I/II devices is highly coupled and dependent on each other.

The aforementioned works all optimize their objectives under the assumption of real-time NSI.
It is generally impractical due to systems' stochasticity and unpredictability, partial feedback, and non-negligible transmission delay.
In \cite{bibli:channel} and \cite{bibli:outdated}, new analytic frameworks were proposed and applied to accommodate outdated NSI, which were able to diminish the optimality loss asymptotically.
Chenshan \emph{et al} in~\cite{bibli:Ren} later extended those frameworks to the scenario of Internet of Things with finite device buffers.
Furthermore, a multi-timescale online algorithm was developed in \cite{bibli:timescale}, where the future queue backlogs were approximated by the current backlogs.
However, all these works did not quantize the tradeoff between feedback signaling and performance loss.

\subsection{Contributions}
Distinctively different from the existing approaches, this paper designs an asymptotically optimal scheduling method for heterogeneous IIoT, where WDs with different battery capabilities are powered by WPT.
Our objective is to maximize a time-average system utility considering both throughput and fairness, and enable the optimization to be tolerant to partial outdated NSI.
The key contributions of this paper are as follows:
\begin{itemize}
  \item We develop an energy-aware resource scheduling (ERS) algorithm to tackle time coupling among data collection, energy transmission, and data offloading in IIoT by leveraging Lyapunov technique.
      A stochastic optimization problem is formulated and then transformed into a series of convex issues with the aid of the data backlog analysis from both the AP and WDs.
      Moreover, we theoretically prove that the proposed algorithm achieves close-to-optimal performance with a known deviation, presenting an [$\mathcal{O}(V)$,$\mathcal{O}(1/V)$]-tradeoff between utility and stability.
      \item
      The proposed ERS algorithm is favorable for multiple resource scheduling in IIoT by exploiting the diversity of heterogeneous devices.
      The objective functions of scheduling Type-I and Type-II devices are meticulously formulated into a similar form and are jointly optimized subject to data stability and energy sustainability.
  \item
      We extend the proposed algorithm to a more practical scenario where only outdated NSI is available for scheduling optimization due to partial feedback and transmission delay.
      Besides, we derive the optimal decisions for the scheduling process and analyze its optimality loss in the presence of stale NSI.
\end{itemize}

Extensive simulations verify the theoretical analyses of the proposed online algorithm for heterogeneous IIoT.
It is shown that the proposed approach, with only partial outdated NSI, can increase the system throughput while maintaining fairness among Type-I/II devices.
Simulations also reveal the impact of feedback overhead and interval on throughput, thereby providing guidelines for the design of compulsory feedback interval in practical implementation.

Note that the selection of Lyapunov optimization techniques in this paper is to design asymptotically optimal and low-complexity resource scheduling approach, which is of practical importance since IIoT requires efficient resource utilization and quick response.
Apart from Lyapunov optimization, the heuristic, game-theoretic, submodular as well as reinforcement learning-based approaches can also be applied to solve the stochastic optimization problem.
However, all these approaches can result in optimality loss due to myopic schedule (e.g., heuristic method) or high complexity and learning time (e.g., Q-learning).
As a result, in the proposed algorithm, we first leverage Lyapunov optimization to decouple the time- and space- couplings in scheduling decisions, and then efficiently solve the decoupled sub-problems by applying convex optimization techniques.

The rest of this paper is organized as follows.
The next section introduces the system model of the considered IIoT and the problem formulation.
Section III proposes the online scheduling algorithm and analyzes its performance theoretically.
Section IV provides the extension of the online algorithm for a more practical scenario.
Simulation results are presented in Section V, followed by the conclusion in Section VI.

\section{System Model and Problem Formulation}\label{section:SM}
\subsection{System Overview}
\begin{figure}
\centering
\subfigure[System scenario]{
\label{fig:scenario} 
\includegraphics[width=0.95\linewidth]{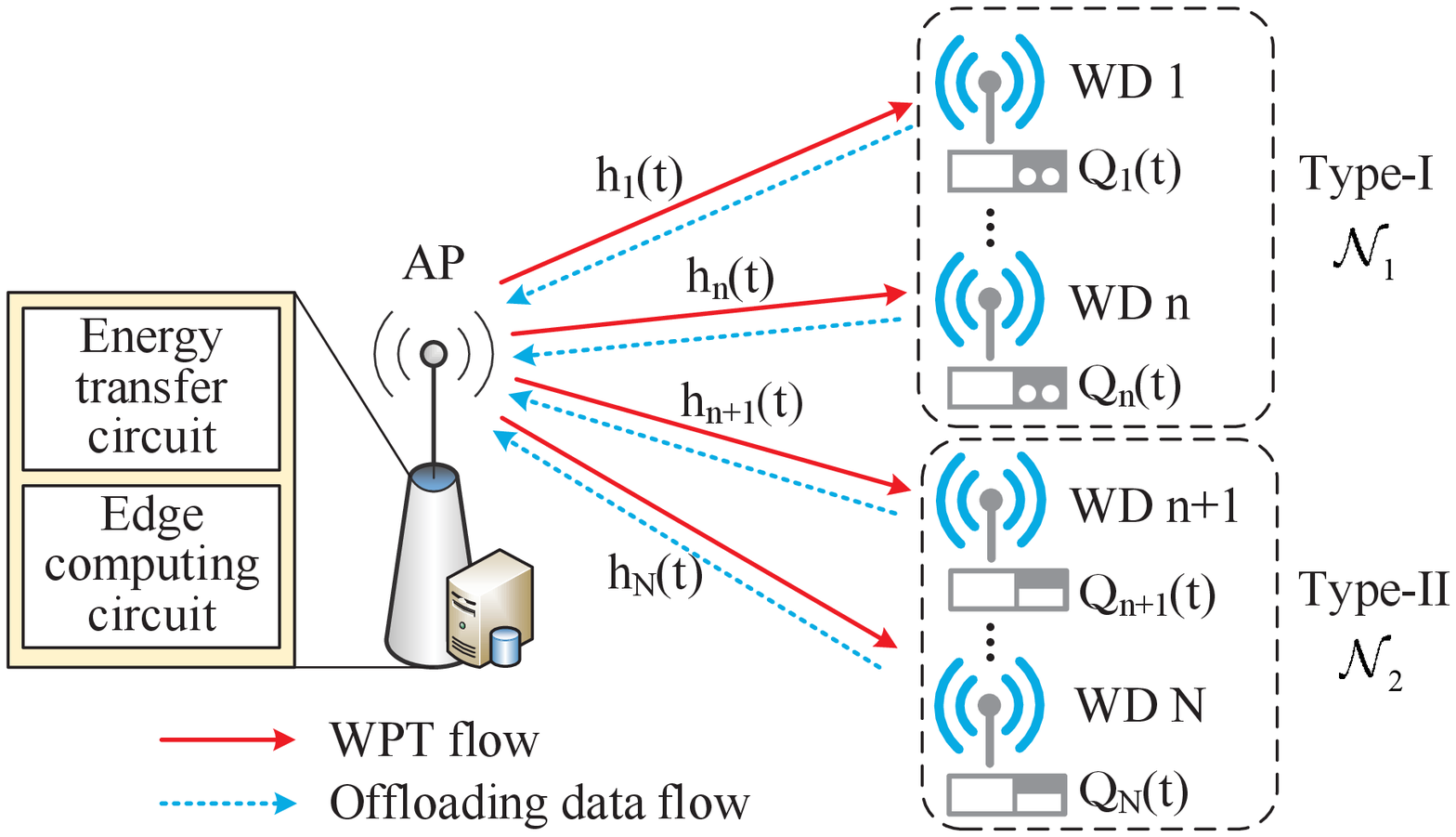}}
\subfigure[Time frame structure]{
\label{fig:protocol}
\includegraphics[width=0.95\linewidth]{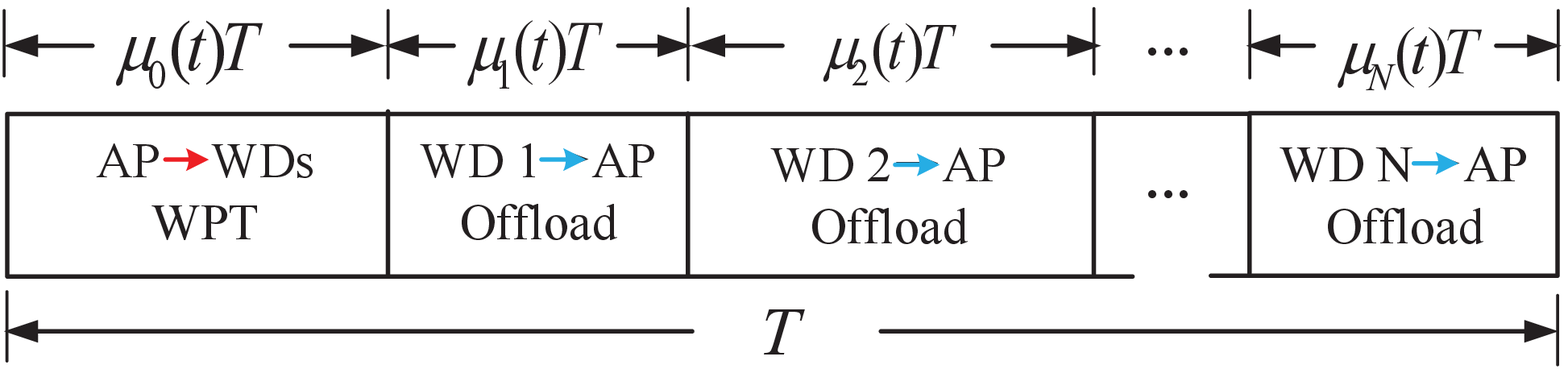}}
\caption{Considered system model of wireless powered MEC for warehouse environmental monitoring in IIoT}
\label{fig:IIoT} 
\end{figure}
We consider an IIoT system consisting of an AP endowed with MEC and WPT functionalities, and $N$ heterogeneous WDs indexed by $\mathcal{N}\!=\!\{1,2,\cdots,N\}$.
As shown in Fig. \ref{fig:scenario}, WDs in the system are divided into sets of Type-I and Type-II devices, denoted by $\mathcal{N}_1$ and $\mathcal{N}_2$, respectively.
Similar to WPCN, we assume the system operates in a frame-based time-division-multiplexing manner\cite{bibli:protocol2,bibli:noise1,bibli:noise2,bibli:CSI}.
Fig. \ref{fig:protocol} illustrates the system' time frame structure with duration $T$, which consists of two phases.
In the first phase, the first $\mu_0(t)T$ amount of time is assigned for downlink WPT, where $\mu_0(t) \in [0,1)$ denotes its allocated time portion at slot $t$.
In the second phase, the rest $(1-\mu_0(t))T$ amount of time is used for the computation offloading in the uplink.
The time allocated for WD $i$ to offload is denoted by $\mu_i(t)T$, where $\mu_i(t) \in [0,1)$.
Consider ambient data analysis for IIoT, the computation results at the AP is not required to be returned to WDs.
Hence, the time portions $\tilde{\bm{\mu}}(t)\!=\!(\mu_i(t),i\in\mathcal{\tilde{N}})$ satisfies
\begin{equation}
\sum\limits_{i \in \mathcal{\tilde{N}}} \mu_i(t) \leq 1,~\mu_i(t)\geq 0,~\forall i \in \mathcal{\tilde{N}},
\vspace*{-3pt}
\label{equ:mu}
\end{equation}
where $\mathcal{\tilde{N}} \triangleq \{0\}\cup\mathcal{N}$ represents the set of the AP and all WDs.

We consider that the AP has a reliable power supply and a constant transmit power $P_0$ by connecting with the power grid.
It is assumed that the AP and all WDs have one single antenna each.
Compared to $P_0$, the energy harvested from noises and received uplink signals from other WDs is much smaller, thus is assumed to be negligible \cite{bibli:protocol2,bibli:noise1,bibli:noise2,bibli:CSI}.
Within time slot $t$, the amount of energy harvested by the $i$-th WD can be expressed as \cite{bibli:protocol2,bibli:delay3}
\begin{equation}
e_i^H(t) = \xi_i P_0 h_i(t) \mu_0(t) T, \ \  i \in \mathcal{N},
\vspace*{-3pt}
\label{equ:energy}
\end{equation}
where $\xi_i \in (0,1)$ is the energy harvesting efficiency, and $h_i(t)$ is the channel power gain between the AP and the $i$-th WD.
Without loss of generality, we assume $e_i^H(t)\!\leq\!e_{i,\text{max}}^H$, where $e_{i,\text{max}}^H$ is the maximum harvested energy in one slot recorded from a long time.
It is further assumed that all channels are independent and identically distributed (i.i.d.) flat block fading, i.e., channels remain static within each time slot but may vary across different slots \cite{bibli:delay,bibli:channel,bibli:measuremnt}.

By exploiting the channel reciprocity in time-division mode system~\cite{bibli:protocol2}~\cite{bibli:delay3}, the AP can acquire both the uplink and downlink channel states at the beginning of each time slot with little signaling cost.
This can be achieved by each WD sending its pilot for channel measurement at the AP~\cite{bibli:CSI,bibli:measuremnt}.
Hence, the achievable offloading data size of WD $i$ (in bits) within time slot $t$ can be given by~\cite{bibli:WMEC1,bibli:delay3}
\begin{equation}
c_i(t) = \mu_i(t) T W \log_{2}(1+\frac{P_i(t)h_i(t)}{N_0}  ),~i \in \mathcal{N},
\vspace*{-3pt}
\label{equ:capacity1}
\end{equation}
where $W$ denotes the communication bandwidth, $P_i(t)$ represents the transmit power of WD $i$, and $N_0$ denotes the noise power at the receiver of the AP.
Given the limited uplink transmission power, $c_i(t)$ in (\ref{equ:capacity1}) is upper bounded by $c_i^{\text{max}}$.
The notations used in this paper are summarized in Table I.
\newcommand{\tabincell}[2]{
}
\begin{table}[!t]
  \centering
  \scriptsize
  \caption{Summary of Notations}
  \label{tab:notations}
  \begin{tabular}{ll}
    \\[-2mm]
    \hline
    \hline\\[-2mm]
    {\bf \small Notation}& {\bf\small Description}\\
    \hline
    \vspace{1mm}\\[-3mm]
    $\mathcal{N} (\mathcal{N}_1,\mathcal{N}_2)$    &  Set of all (Type-I, Type-II) wireless devices \\
    \vspace{1mm}
    $\xi_i$   &  Energy harvesting efficiency of device $i$ \\
    \vspace{1mm}
    $P_0$   &  Transmit power of AP \\
    \vspace{1mm}
    $P_i(t)$   &  Offloading transmit power of device $i$ at slot t\\
    \vspace{1mm}
    $h_i(t)$   &  Channel power gain between AP and device $i$ at slot t \\
    \vspace{1mm}
    $T$      &  Interval of a time block \\
    \vspace{1mm}
    $W$      &  System bandwidth \\
    \vspace{1mm}
    $\mu_0(t)$  &  Time portion for wireless power transfer at slot $t$\\
    \vspace{1mm}
    $\mu_i(t)$  &  Time portion for device $i$ to offload at slot $t$\\
     \vspace{1mm}
    $c_i(t)$    & Achievable offloading data size of device $i$ at slot $t$ \\
    \vspace{1mm}
     $E_i(t)$  &   Available energy of device $i$ at slot $t$\\
     \vspace{1mm}
     $e_i(t)$  &   Offloading transmit energy of device $i$ at slot $t$\\
     \vspace{1mm}
     $e_i^H(t)$  &   Energy harvested by device $i$ at slot $t$\\
     \vspace{1mm}
    $A_i(t)$  &   Data that can be collected by device $i$ within slot $t$\\
     \vspace{1mm}
    $a_i(t)$  &   Data collected at device $i$ within slot $t$\\
    \vspace{1mm}
    $Q_i(t)$  &   Data queue backlog of device $i$ at slot $t$\\
    \vspace{1mm}
    $S_i(t)$  &   Data queue backlog of AP at slot $t$\\
     \vspace{1mm}
    $r_i(t)$  &   Available data processing speed of AP at slot $t$\\
     \vspace{1mm}
    $\overline{X}$  &   Time-average of any stochastic process $X$\\
     \vspace{1mm}
    $U_i(\cdot)$  &   Utility of device $i$\\
    \hline
    \hline
  \end{tabular}
\end{table}

\subsection{Energy Models for Heterogeneous Devices}
Energy models for Type-I and Type-II devices are entirely different due to their battery management capabilities, which are further discussed in the following.
\subsubsection{Type-I devices}
This type of WDs is equipped with batteries that have low energy storage capacity and high self-discharge rate.
In this case, the harvested but unused energy within the current slot will be depleted and cannot be kept for future use.
As a result, if $\mu_i(t)\!\neq\!0$, $i$-th WD will manage its energy in a myopic manner, i.e., exhausting all the harvested energy for offloading.
Thus, we have $P_i(t)\!=\!0$ if WD $i$ stay silent in the offloading phase, otherwise, WD $i$ will offload with transmit power
\begin{equation}
P_i(t) = \frac{\eta_ie_i^H(t)}{\mu_i(t)T},~i \in \mathcal{N}_1,
\vspace*{-3pt}
\label{equ:power}
\end{equation}
where $\eta_i$ represents the fixed portion of the harvested energy that is used for data offloading.
As the device exhausts all the harvested energy for uplink transmission, $\eta_i\!=\!1$ in the sequel.
The maximum transmit power constraint for Type-I devices is neglected here since the harvested energy from WPT at each slot is small in practice \cite{bibli:delay3}.

\subsubsection{Type-II devices}
Different from Type-I devices, each device $i \in \mathcal{N}_2$ is equipped with a rechargeable battery that has a high energy storage capacity and low discharge rate.
As long as the maximum battery capacity $\theta_i$ is not exceeded, a Type-II device can save the harvested energy in its battery for offloading in the following slots when the channel fading condition is unfavorable.
Denote $\bm{E}(t)\!=\!(E_i(t),i \in \mathcal{N}_2)$ as the vector of Type-II devices' energy queue sizes at slot $t$, and $\bm{e}(t)\!=\!(e_i(t),i\in\mathcal{N}_2)$ as the vector of their corresponding uplink transmit energy.
The available energy in WD $i$ evolves according to the dynamic equation
\begin{equation}
E_i(t+1) = \min \{E_i(t) + e_i^H(t), \theta_i \} - e_i(t),~i \in \mathcal{N}_2.
\label{equ:energy-B}
\end{equation}
At any time slot $t$, the transmit energy $e_i(t)$ must follow the \emph{energy-available} condition
\begin{equation}
e_i(t) \leq E_i(t)+e_i^H(t),~i \in \mathcal{N}_2,
\label{equ:battery1}
\end{equation}
such that the energy used would not exceed what has been cumulatively harvested so far.
Meanwhile, since a Type-II device can accumulate much energy before its transmission, when offloading, it should also satisfy the maximum transmit power constraint
\begin{equation}
e_i(t) \leq P_i^{\text{max}}\mu_i(t)T,~i \in \mathcal{N}_2,
\label{equ:battery2}
\end{equation}
where $P_i^{\text{max}}$ is $i$-th WD's maximum transmit power.

\subsection{Data Collection and Queueing Models}
At any slot $t$, the volume of data that can be collected by WD $i$ is denoted by $A_i(t)$.
Due to the dynamic changes of environment in IIoT, we assume $A_i(t)$ to be i.i.d. over slots with a maximum $A_i^{\text{max}}$~\cite{bibli:arrival,bibli:channel}.
Given the data freshness and the limited data buffer of WD, only part of the data, denoted by $a_i(t)$, would be collected into the data buffer.
Hence, we have
\begin{equation}
0 \leq a_i(t) \leq A_i(t)\leq A_i^{\text{max}}, ~\forall i \in \mathcal{N}.
\label{equ:a}
\end{equation}
Let $\bm{Q}(t)\!=\!(Q_i(t),i \in \mathcal{N})$ denote the backlog of data queue of WDs at slot $t$, i.e., any collected but have not been offloaded data that are queued in the data buffer.
At every slot, the network decides how much data should be collected (i.e., $a_i(t)$) and how much existing data in the buffer to offload (i.e., $c_i(t)$) according to the network state.
Follow the basic operations of cascaded queueing systems~\cite{bibli:arrival,bibli:channel} (e.g., the edge computing network where the data arrival and offloading operate on a first-in-first-out basis), the amount of data to be collected (i.e., $a_i(t)$) is determined at slot $t$ and queued into the data buffer at the beginning of slot $t\!+\!1$.
Therefore, the data backlog $Q_i(t)$ evolves as follows:

\begin{equation}
Q_i(t+1)\!=\! [Q_i(t) - c_i(t)]^+ + a_i(t),~i \in \mathcal{N},
\label{equ:Qt}
\end{equation}
where $[x]^+\!=\!\max\{x,0\}$.
In some special cases, for example, at the early stage of the network, $Q_i(t)$ may be smaller than $c_i(t)$ as the collection of data takes time. Here, the first term in the right-hand-side (RHS) of equation (\ref{equ:Qt}) is used to guarantee that the device cannot offload more than what remains in its data buffer when $Q_i(t)\!<\!c_i(t)$ happens~\cite{bibli:arrival,bibli:dummy1}.

The AP maintains $N$ data buffers $\bm{S}(t)\!=\!(S_i(t),i \in \mathcal{N})$ to store the data that offloaded from WDs but have not been processed.
Constrained by the limited computing capability of the AP, at most $r_i(t)$ amount of data from $i$-th WD can be processed at slot $t$, where $r_i(t)$ is a stochastic number with the maximum $r_i^{\text{max}}$ \cite{bibli:channel}.
Naturally, the data queue dynamics at the AP are given by
\begin{equation}
S_i(t+1) = [S_i(t) - r_i(t)]^+ + \min\{c_i(t),Q_i(t)\},
\label{equ:Ft}
\end{equation}
where the second term in the RHS of (\ref{equ:Ft}) represents the \emph{data-available} constraint of WD $i$, which means the WD cannot offload more data than what it has stored.
In the case when $c_i(t)\!>\!Q_i(t)$, the excessive offloading rate will be used for transmitting dummy data~\cite{bibli:arrival,bibli:dummy1}.
We will later show that the optimization is still asymptotically optimal by doing this.

\subsection{Problem Formulation}
Consider warehouse environmental monitoring services in an IIoT system, unfair data collection and offloading would reduce the overall system performance.
This is because we may waste too many resources for offloading redundant data from one place while lacking enough data from other areas for analysis.
Therefore, in this paper, we aim at maximizing the system throughput while guaranteeing fairness among WDs.

However, fairness in the considered scenario is not easy to achieve due to the \emph{doubly near-far} phenomenon \cite{bibli:protocol2} and heterogeneity of the WDs, which is further discussed in section~\ref{section:sim}.
To address this issue, instead of directly maximizing the throughput, we formulate the problem as a sum-utility maximization problem as follows
\begin{equation*}
\begin{aligned}
\textbf{P:}~~&\mathop{\max}_{\tilde{\bm{\mu}}(t),\bm{e}(t),\bm{a}(t)}~
\sum\limits_{i \in \mathcal{N}}~\overline{U_i(a_i(t))}
\\ &~~~~~~\text{s.t.} ~~(\ref{equ:mu}),(\ref{equ:battery1}),(\ref{equ:battery2}),(\ref{equ:a}),\\
&~~~~~~~~~~~\textbf{C1:}~\overline{Q_i}< \infty,~\forall i \in \mathcal{N},\\
&~~~~~~~~~~~\textbf{C2:}~\overline{S_i}< \infty,~\forall i \in \mathcal{N},
\end{aligned}
\label{equ:utility0}
\end{equation*}
where $\bm{a}(t)\!=\!(a_i(t),i \in \mathcal{N})$, $\overline{X}\!=\!\lim_{t \to \infty} \frac{1}{t} \sum\nolimits_{\tau =0}^{t-1} \mathbb{E}\{ X(\tau) \}$ defines the time average expectation of any stochastic process $X(t)$, constraints \textbf{C1} and \textbf{C2} ensure the stability of all the data queues, and the utility function based on proportional fairness~\cite{bibli:fairness,bibli:channel} is defined as $U_i(x)\!=\!\log(1\!+\!x)$.
Here, $\log(\cdot)$ denotes the natural logarithm.

As can be seen from the utility function, the marginal utility decreases as the amount of data collected by a WD increase.
As a result, the system would be inclined to collect equal amounts of data from different WDs rather than show an apparent preference for some devices.
This thereby ensures system fairness.

The problem \textbf{P} is a typical stochastic optimization problem.
An offline optimization of \textbf{P} would require complete non-causal NSI, which is impossible in practice due to the highly stochastic and unpredictable characteristics of IIoT systems.
Even if such NSI is available, the problem is still challenging due to its high computational complexity.
Therefore, we are motivated to propose online algorithm to jointly schedule $\tilde{\bm{\mu}}(t)$, $\bm{e}(t)$ and $\bm{a}(t)$ without future NSI.

\section{Online Schedule of Heterogeneous Devices}\label{section:proposed}
In this section, we develop an online decision-making algorithm by employing Lyapunov optimization.
The original problem \textbf{P} is decoupled into a series of deterministic per-time slot problems at independent time slots, which are then solved by convex optimization technique.
We later prove that the proposed algorithm preserves asymptotic optimality compared with the optimum of the original problem.

\subsection{Problem Transformation}
Define $\bm{\Theta}(t)\! =\![\bm{Q}(t),\bm{E}(t),\bm{S}(t)]$ as the concatenated vector of the queues at WDs and the AP.
To handle the time-coupling in problem~$\textbf{P}$, we first define a non-negative perturbed Lyapunov function as
\begin{equation}
L\big(\bm{\Theta}(t)\big)\!=\! \frac{1}{2}\left\{ \sum\limits_{\mathclap{{i \in \mathcal{N}} }}S_i(t)^2\!+\! \sum\limits_{\mathclap{ {i \in \mathcal{N}} }}Q_i(t)^2\!+\! \sum\limits_{\mathclap{ {i \in \mathcal{N}_B} }}(E_i(t)-\theta_i)^2 \right\},
\end{equation}
where $\theta_i$ is a weight perturbation, also the battery capacity of Type-II devices mentioned in Section \ref{section:SM}.
We will later show that (\ref{equ:battery1}) can be satisfied with a proper choice of $\theta_i$.
Meanwhile, when minimizing $L\big(\bm{\Theta}(t)\big)$, we push the data backlogs at both the AP and WDs towards zero, which is equivalent to satisfying stability constraints \textbf{C1} and \textbf{C2} in \textbf{P}.
Then the one-slot conditional Lyapunov drift can be expressed as
\begin{equation}
\Delta(\bm{\Theta}(t))\!=\! \mathbb{E}\big[L\big(\bm{\Theta}(t+1)\big)-L\big(\bm{\Theta}(t)\big)\mid\bm{\Theta}(t)\big].
\label{equ:drift}
\end{equation}
where the expectation is taken with respect to the random NSI and the control actions.
Based on Lyapunov optimization technique, we consider minimizing the drift-plus-penalty expression
\begin{equation}
\Delta_V\big(\bm{\Theta}(t)\big)\!=\!\Delta\big(\bm{\Theta}(t)\big)\!-\! V\mathbb{E}\left\{\sum\limits_{i \in \mathcal{N}} ~U_i\big(a_i(t)\big)\mid \bm{\Theta}(t) \! \right\},
\end{equation}
instead of directly leveraging the objection in \textbf{P}.

\renewcommand{\IEEEQED}{\IEEEQEDopen}
\newtheorem{lemma}{Lemma}
\begin{lemma}
\label{lemma:bound}
For any optimization decisions made on slot~$t$, and all possible values of $\bm{\Theta}(t)$, the drift-plus-penalty expression for all slot $t$ satisfies
\begin{equation}
\begin{aligned}
\Delta_V\big(\bm{\Theta}(t)\big)
\leq & B_1 -V \mathbb{E}\left\{\sum\limits_{i \in \mathcal{N}}~U_i\big(a_i(t)\big)\mid \bm{\Theta}(t) \! \right\}\\
&~+\sum\limits_{i \in \mathcal{N}} Q_i(t)\mathbb{E}\left\{ a_i(t)-c_i(t) \mid \bm{\Theta}(t)  \right\}\\
&~+\sum\limits_{i \in \mathcal{N}} S_i(t)\mathbb{E}\left\{ c_i(t)-r_i(t) \mid \bm{\Theta}(t)  \right\}\\
&~+\sum\limits_{i \in \mathcal{N}_2} \left[E_i(t)\!-\!\theta_i \right]\mathbb{E}\{e_i^H(t)\!-\!e_i(t)\mid \bm{\Theta}(t)\},
\end{aligned}
\label{equ:d-p-p}
\end{equation}
where $V \!>\! 0$ is a control parameter that affects an explicit tradeoff between system utility and data backlogs, $B_1$ denotes a finite constant that satisfies
\begin{equation}
\begin{aligned}
B_1\!&=\! \frac{1}{2}\left\{ \sum\limits_{i \in \mathcal{N}}\left[ (A_i^{\text{max}})^2+ (r_i^{\text{max}})^2\right]\!+\!2\sum\limits_{i \in \mathcal{N}}(c_i^{\text{max}})^2\right\}\\
&+\frac{1}{2}\left\{\sum\limits_{i \in \mathcal{N}_2}[(e_{i,\text{max}}^H)^2+(P_i^{\text{max}}T)^2]\right\}
\label{equ:bound}
\end{aligned}
\end{equation}
\end{lemma}

\textit{Proof:}~Please refer to Appendix A

Minimizing $\Delta_V\big(\bm{\Theta}(t)\big)$ is still difficult due to its dynamics.
Instead, we are motivated to minimize the RHS of (\ref{equ:d-p-p}) according to the principle of opportunistically minimizing an expectation \cite{bibli:Lyapunov} in a per-slot manner, subject to the constraints (\ref{equ:mu}),~(\ref{equ:battery2}), and (\ref{equ:a}).

\subsection{Optimal Scheduling Decisions}
In this subsection, we present an \textbf{E}nergy-aware \textbf{R}esource \textbf{S}cheduling algorithm with \textbf{R}eal-time \textbf{N}SI (ERS-RN algorithm) to address the above problem.
Since $\bm{a}(t)$ and $(\bm{e}(t),\tilde{\bm{\mu}}(t))$ can be decoupled with each other and are independent of the current backlog $\bm{\Theta}(t)$, the minimization problem can be separated into two optimization sub-problems as follows.

\subsubsection{\textbf{Distributed Data Collection}}
For each time slot $t$, the data collection sub-problem can be decomposed for individual WD.
The optimal data collection strategy can be obtained by solving
\begin{equation*}
\begin{aligned}
\textbf{P1:} ~~~~\mathop{\min}_{a_i(t)}~~&
Q_i(t)a_i(t) - VU_i\big(a_i(t)\big)\\
\text{s.t.} ~~~&(\ref{equ:a}).
\end{aligned}
\label{equ:P2-1}
\end{equation*}
The problem \textbf{P1} is a convex optimization problem since the objective function is convex and its constraint is linear.
Hence, the optimum is either at the stationary point of $a_i(t)=V/Q_i(t)\!-\!1$ or on one of the boundaries.
For any WD $i \in \mathcal{N}$, the optimal data collection strategy is thus chosen by
\begin{equation}
a_i(t)=
\begin{cases}
A_i(t),& \text{$V \geq (A_i(t)+1)Q_i(t)$}\\
[\frac{V}{Q_i(t)}-1]^+,& \text{otherwise}
\end{cases}
\label{equ:condition1}
\end{equation}

The given policy indicates that the amount of data collected by each WD depends on the length of its backlog.
A device is inclined to collect more data when its backlog is small.

\subsubsection{\textbf{Joint Energy and Time Allocation}}
According to (\ref{equ:d-p-p}), the optimal transmit powers $\bm{e}(t)$ and the optimal time portions $\tilde{\bm{\mu}}(t)$ can be derived by solving
\begin{equation*}
\begin{aligned}
\textbf{P2:} \mathop{\min}_{\bm{e}(t),\tilde{\bm{\mu}}(t)}&~
\sum\limits_{\mathclap{{i \in \mathcal{N}} }}[S_i(t)-Q_i(t)]c_i(t)\\
&\!+\!\sum\limits_{\mathclap{{i \in \mathcal{N}_2} }}[E_i(t)\!-\!\theta_i][e_i^H(t)\!-\!e_i(t)]\\
\text{s.t.} ~~~&(\ref{equ:mu}), (\ref{equ:battery2}).
\end{aligned}
\label{equ:P2-4}
\end{equation*}

Due to the constraint (\ref{equ:mu}), $\bm{e}(t)$ and $\tilde{\bm{\mu}}(t)$ in problem \textbf{P2} cannot be completely decoupled.
Generally, \textbf{P2} is a non-convex optimization problem.
To solve this problem, we first define a set $\mathcal{N}_t \triangleq \{i\mid i\!\in\! \mathcal{N}, S_i(t)\! \geq\! Q_i(t)\}$, and then find an interesting property of its optimization as shown in the following theorem.

\newtheorem{theorem}{Theorem}
\begin{theorem}
\label{lemma:SP}
For any device $i \in \mathcal{N}_t \cap \mathcal{N}_2$, the term $[S_i(t)-Q_i(t)]c_i(t)+[E_i(t)\!-\!\theta_i][e_i^H(t)\!-\!e_i(t)]$ in the objective function of \textbf{P2} gets its minimum when $\mu_i(t)=0,~e_i(t)=0$.
Besides, for arbitrary WD $i \in \mathcal{N}_t \cap \mathcal{N}_1$, its optimal time portion is given by $\mu_i(t)=0$.
\end{theorem}

\textit{Proof:}~Please refer to  Appendix \ref{section:zero}

\newtheorem{remark}{Remark}
\begin{remark}
Theorem \ref{lemma:SP} suggests that the AP prefers to serve WDs who have a larger data backlog than that queueing at the AP.
This result is in accordance with our intuition that when the waiting queue at the AP is large, offloading data from a WD whose queue length is small only brings transmission energy loss and more execution delay.
\end{remark}

By excluding the devices in $\mathcal{N}_t$, we denote the remaining Type-I WDs to be scheduled as $\mathcal{N}_1^t$.
Similarly, let $\mathcal{N}_2^t$ collect the Type-II devices that are not in $\mathcal{N}_t$.
For the sake of brevity, we define $D_i(t)\!=\![S_i(t)\!-\!Q_i(t)]TW$, $\beta_i(t)\!=\!h_i(t)/(N_0T)$, $\delta_i(t)\!=\!\xi_i P_0 h_i^2(t)/N_0$ in the sequel.
Then the problem \textbf{P2} can be simplified and rewritten as follows
\begin{equation*}
\begin{aligned}
\textbf{P3:}~~&\mathop{\min}_{\bm{v}(t)}~~
\sum\limits_{\mathclap{{i \in \mathcal{N}_2^t} }}D_i(t)\phi_i(t)\!-\sum\limits_{i \in \mathcal{N}_2^t}[E_i(t)\!-\!\theta_i]e_i(t)\\
&\!+\!\sum\limits_{\mathclap{{i \in \mathcal{N}_1^t} }}D_i(t)\psi_i(t)\!+\!\mu_0(t)TP_0\sum\limits_{\mathclap{ {i \in \mathcal{N}_2} }}[E_i(t)\!-\!\theta_i]\xi_i h_i(t)\\
&~\text{s.t.} ~\textbf{C3:}\sum\limits_{\mathclap{{i \in \tilde{\mathcal{N}} \backslash \mathcal{N}_t} }} \mu_i(t) \leq 1,~\mu_i(t) > 0,~\forall i \in \tilde{\mathcal{N}} \backslash \mathcal{N}_t,
\\&~~~~~\textbf{C4:}~e_i(t) \leq P_i^{\text{max}}\mu_i(t)T,~i \in \mathcal{N}_2^t,
\end{aligned}
\end{equation*}
where $\bm{v}(t)\!=\!\{e_i(t)|i \in \mathcal{N}_2^t\}\cup \{\mu_i(t)|i \in \tilde{\mathcal{N}} \backslash \mathcal{N}_t \}$ collects all variables that is to be optimized, and
\begin{equation}
\phi_i(t)\!=\!\mu_i(t) \log_{2}(1\!+\!\frac{\beta_i(t)e_i(t)}{\mu_i(t)}),~i \in \mathcal{N}_2^t,
\label{equ:phi}
\end{equation}
\begin{equation}
\psi_i(t)=\mu_i(t) \log_{2}(1+\frac{\delta_i(t)\mu_0(t)}{\mu_i(t)}),~i \in \mathcal{N}_1^t.
\label{equ:psi}
\end{equation}

Note from (\ref{equ:phi}) that $\phi_i(t)$ is a perspective function of $f(x)=\log_{2}(1+\beta_i(t) x)$.
Since $f(x)$ is a concave function of $x \geq 0$, $\phi_i(t)$ is jointly concave with respect to $e_i(t)\! >\! 0$ and $\mu_i(t)\! >\! 0$ for arbitrary $i \in \mathcal{N}_2^t$ \cite{bibli:convex}.
Similarly, we can prove that $\psi_i(t)$ is concave in \{$\mu_0(t),\mu_i(t)\!:\!\forall i\in \mathcal{N}_1^t$\} when $\mu_0(t)\! >\! 0$ and $\mu_i(t)\! >\! 0$.
Since the objective function in \textbf{P3} is a negative weighted
sum of $\phi_i(t)$, $\psi_i(t)$ and several linear functions, the objective function of \textbf{P3} is convex in $\bm{v}$(t).
Meanwhile, the constraints \textbf{C3} and \textbf{C4} are linear, and thus convex.
Therefore, the problem \textbf{P3} is a convex optimization problem, which can be effectively solved by off-the-shelf toolbox CVX \cite{bibli:CVX} at a maximal computational complexity order of $\mathcal{O}(\max\{(2|\mathcal{N}_2^t|\!+\!|\mathcal{N}_1^t|\!+\!1)^3,C\})$\cite{bibli:Lyapunov}.
Here, $(2|\mathcal{N}_2^t|\!+\!|\mathcal{N}_1^t|\!+\!1)$ represents the number of variables to be optimized, $C$ is the cost to evaluate the objective function together with its first and second derivatives.

\begin{algorithm}[t]
\caption{ERS-RN Algorithm}
\label{alg:1}
\LinesNumbered 
\textbf{At the AP}:\\
1:~Observe $\bm{Q}(t)$, $\bm{S}(t)$, $\bm{E}(t)$, $\bm{r}(t)$ and channel gains;\\
2:~Choose $\bm{v}(t)$ by solving \textbf{P3};\\
3:~Inform $\bm{v}(t)$ to WDs and execute WPT;\\
4:~Update $E_i(t)$ at each Type-II device $i$ by $\bm{v}(t)$ and (\ref{equ:energy-B}).\\
\textbf{At any WD $i$}:\\
5:~Observe $Q_i(t)$, $A_i(t)$ and $c_i(t)$;\\
6:~Decide $a_i(t)$ based on (\ref{equ:condition1});\\
7:~Offload data and give feedback $Q_i(t)$ to the AP in the arranged sequence;\\
8:~Update $Q_i(t)$ according to (\ref{equ:Qt}).\\
\textbf{At the AP}:\\
9:~Update $\bm{S}(t)$ according to (\ref{equ:Ft}).
\end{algorithm}

Based on the above analysis, the whole process of the ERS-RN algorithm is summarized in \textbf{Algorithm 1}.
We call the algorithm energy-aware because the AP conducts the energy allocation by itself and obtains the real-time knowledge of all devices' energy storages without any feedback from WDs.
Note that the joint energy and time allocation depends on real-time data queue lengths fed back by all WDs.
Hence, the optimization of time portions can only be coordinated at the AP in a centralized manner.

\subsection{Performance Analysis}
In this subsection, we start by showing that the proposed ERS-RN algorithm is asymptotically optimal.
Denote $U^\star$ as the offline optimum of problem \textbf{P}, which can only be obtained by using full non-causal knowledge of the IIoT network.
Let $U^*$ denote the long-term time-average utility achieved by the ERS-RN algorithm.
Then, we can establish the following theorem.
\begin{theorem}
\label{lemma:O1V}
Suppose the problem \textbf{P} is feasible, the gap between $U^*$ and $U^\star$ satisfies
\begin{equation}
U^\star - U^* \leq B_1/V.
\end{equation}
\end{theorem}

\textit{Proof:}~Please refer to  Appendix \ref{section:O1V}

Theorem \ref{lemma:O1V} suggests that the achievable time-average utility of the proposed method can be arbitrarily close to the offline optimum by setting a sufficiently large $V$.
In other words, the proposed approach is asymptotically optimal in terms of solving the original problem \textbf{P}.
However, it is not the larger the value of $V$, the better.
The increase of $V$ would also cause a negative impact on system data backlogs, as revealed in the following theorem.

\begin{theorem}
\label{lemma:OV}
At any time slot, the data backlogs at WD $i \in \mathcal{N}$ and the AP are all upper bounded, with upper bounds given by
\begin{equation}
Q_i^{\text{max}}\! =\! V\!+\!A_i^{\text{max}},
\label{equ:Qmax}
\end{equation}
\begin{equation}
S_i^{\text{max}}\! =\! V\!+\!A_i^{\text{max}}\!+\!c_i^{\text{max}}.
\label{equ:Zmax}
\end{equation}
\end{theorem}

\textit{Proof:}~Please refer to  Appendix \ref{section:OV}

With Theorem \ref{lemma:OV}, we then provide the value of the weight perturbation parameter $\theta_i$ and prove the \emph{energy-availability} constraint (\ref{equ:battery1}) is always satisfies in the following theorem.
\begin{theorem}
\label{lemma:theta}
With the proposed algorithm, the energy storage capacity (or the perturbation parameter) $\theta_i$ is given as follows
\begin{equation}
\theta_i = \frac{(V+A_i^{\text{max}})c_i^{\text{max}}}{e_i^{\text{min}}}+P_i^{\text{max}}T,
\label{equ:theta}
\end{equation}
where $e_i^{\text{min}}$ is the minimum transmit energy other than zero.
The equation (\ref{equ:theta}) ensures there is enough energy in the battery for optimal schedule. In other words, $E_i(t)+e_i^H(t)>P_i(t)T$.
\end{theorem}

\textit{Proof:}~Please refer to  Appendix \ref{section:energy}

Theorem \ref{lemma:OV} and \ref{lemma:theta} show that $Q_i^{\text{max}}$, $S_i^{\text{max}}$ and $\theta_i$ increase monotonically with the control parameter $V$.
A larger $V$ leads to a higher system utility at the expense of larger queue lengths at all WDs and the AP.
On the other hand, a tiny $V$ impairs the asymptotic optimality of achievable utility but reduces the demands for large storage spaces of the system.
This reveals a $[O(1/V),O(V)]$ tradeoff between the system utility and the required storage spaces.
Therefore, instead of giving each WD a substantial data and energy storage space, we can reduce the manufacturing costs of IIoT systems by embedding appropriate data buffers and batteries for WDs based on the expected system utility.

\section{Resource Scheduling in IIoT with Implementation Considerations}
In this section, we consider a more general setting in which the overhead and delay for transmitting NSI cannot be overlooked.
We develop a modified version of the ERS-RN algorithm, called ERS-ON, to address the issues brought by practical application, and further analyze its performance theoretically.

\subsection{Practicability Considerations}
As discussed, real-time NSI feedback (i.e., $\bm{Q}(t)$) is needed at the AP for the joint optimization of energy allocations and time portions.
However, the transmission of NSI would occupy the time for data offloading, which may jeopardize network performance, or even prohibit the system from working.
To avoid explosive feedback costs, systems may send back only partial NSI, as the majority of feedbacks have less contribution for optimization.
Moreover, even if all WDs give continuous feedbacks at every slot, the AP can only get outdated NSI (e.g., $\bm{Q}(t\!-\!1)$ at slot $t$) since it has to make the scheduling decision before obtaining $Q_i(t)$ in the uplink phase at slot $t$.
As a result, the proposed ERS-RN algorithm may be hard to be directly applied to real systems.
Some modifications are needed in the presence of outdated NSI.

Therefore, in this section, we proceed to consider a more practical scenario of the considered IIoT system, where any device~$i$ that has been assigned time for offloading at slot~$t$ should spare at least $\epsilon_i T$ time to give feedback of its data backlog.
Here, the time portion $\epsilon_i$ is a constant obtained by historical experience such that there is always enough energy for transmitting an $L$ bits feedback.
WDs that have not gotten any uplink chances (i.e., silent) at slot $t$ are generally not required for feedbacks in order to reduce system NSI overheads.
The only exception is for those WDs that remain silent in the past $m$ slots, in case the shortage of NSI update at the AP permanently prevents some WDs from offloading.
For notation simplicity, we denote those WDs that reach their compulsory feedback interval $mT$ at time slot $t$ as~$\mathcal{M}_t$ in the sequel.

\subsection{Energy-Aware Resource Scheduling Algorithm with Outdated NSI}
The outdated NSI (i.e., data backlogs at WDs) acquired through the above feedback mechanism has a direct impact on the optimization of problem \textbf{P3}.
To overcome the limitation of the ERS-RN algorithm, we develop a modified version of the ERS algorithm with Outdated NSI (ERS-ON) by approximating the latest feedback of WDs' data backlog values as the current values.
Let $\widehat{Q}_i(t)$ denote the approximation of $i$-th WD's data backlog that the AP holds at time slot $t$.
Hence, we have $\widehat{Q}_i(t)=Q_i(t-\tau_i)$, where $\tau_i \in \{1,\cdots,m\}$.
Then, the problem \textbf{P3} can be transformed into
\begin{equation*}
\begin{aligned}
\textbf{P4:}~~&\mathop{\min}_{\bm{v}(t)}~~
\sum\limits_{\mathclap{{i \in \mathcal{N}_2^t} }}\widehat D_i(t)\phi_i(t)\!-\sum\limits_{i \in \mathcal{N}_2^t}[E_i(t)\!-\!\theta_i]e_i(t)\\
&\!+\!\sum\limits_{\mathclap{{i \in \mathcal{N}_1^t} }}\widehat D_i(t)\psi_i(t)\!+\!\mu_0(t)TP_0\sum\limits_{\mathclap{ {i \in \mathcal{N}_2} }}[E_i(t)\!-\!\theta_i]\xi_i h_i(t)\\
&\text{s.t.} ~\textbf{C4}, \\&~~~~\textbf{C5:}\sum\limits_{\mathclap{{i \in \tilde{\mathcal{N}} \backslash \mathcal{N}_t} }} \mu_i(t) \leq 1\!-\!\sum\limits_{\mathclap{{i \in \mathcal{M}_t }}}\epsilon_i,~\mu_i(t) > \epsilon_i,~\forall i \in \mathcal{N} \backslash \mathcal{N}_t.
\end{aligned}
\end{equation*}
where $\widehat{D}_i(t)\!=\![S_i(t)-\widehat{Q}_i(t)]TW$, and \textbf{C5} ensures there is enough time for offloading and feedback in each time slot. Similar to \textbf{P3}, problem \textbf{P4} can be solved by CVX or other convex optimization techniques as the constraint \textbf{C5} preserves convexity.

Let $\{e_i^*(t)\!\mid i\in \mathcal{N}_2^t\}$ collect the optimal energy solutions and $\{\mu_i^*(t)\mid i\in \tilde{\mathcal{N}} \backslash \mathcal{N}_t \}$ denote the optimal time portions of problem \textbf{P4}.
The time portions for WDs and the AP at time slot $t$ can thus be given by
\begin{equation}
\mu_i(t)=
\begin{cases}
~0,& \text{$i \in \mathcal{N}_t \backslash \mathcal{M}_t$}\\
~\epsilon_i,& \text{$i \in \mathcal{M}_t$}\\
~\mu_i^*(t),& \text{otherwise},
\end{cases}
\end{equation}
where $\mu_i(t)\!=\!0$ means WD~$i$ keeps silent at slot~$t$.

Different from the time portions allocation, the energy policies for WDs vary with their types.
For Type-I devices, same like the statement in section II, all the harvested energy would be consumed for offloading/feedback or be depleted.
On the contrary, Type-II devices in $\mathcal{N}_2 \cap \mathcal{M}_t$ can dynamically adjust their transmit energy to ensure the channel capacity equals the data size of feedback, i.e., $L\!=\!TW\phi_i(t)$, thus saving excessive energy for future use.
More specifically, the energy allocation policy for device $i \in \mathcal{N}_2$ satisfies
\begin{equation}
e_i(t)=
\begin{cases}
~\frac{\epsilon_i}{\beta_i(t)}\big [2^{l_i}-1 \big ],& \text{$i \in \mathcal{N}_2 \cap \mathcal{M}_t$}\\
~e_i^*(t),& \text{$i \in \mathcal{N}_2^t$}\\
~0,& \text{otherwise},
\end{cases}
\end{equation}
where $l_i\!=\!L/(\epsilon_i T\!W)$.
The allocated transmit energy of Type-II devices would not violate the constraints (\ref{equ:battery1}) and (\ref{equ:battery2}) since we have a large $\beta_i(t)$ and small $l_i$ when WD $i \in \mathcal{N}_2 \cap \mathcal{M}_t$.

\subsection{Asymptotic Optimality Analysis under Outdated NSI}
The use of outdated NSI will inevitably impair the performance of systems.
In the worst case, a WD would lose $m$ chances to offload during the past $mT$ time duration due to a lack of information update at the AP.
For any slot $t$, the difference between the approximate values maintained at the AP and the actual data backlogs at WDs are bounded by

\begin{equation}
\begin{aligned}
Q_i(t)-\widehat{Q}_i(t)&\!=\! \sum\limits_{\mathclap{{\tau=t\!-\!\tau_i\!+\!1}}}^{t} \big [Q_i(\tau)-Q_i(\tau-1)\big ]\\
&\!\leq\! \tau_i A_i^{\text{max}}\! \leq \!m A_i^{\text{max}},
\end{aligned}
\label{equ:Q-out-B}
\end{equation}
where the first inequality is derived from (\ref{equ:a}) and (\ref{equ:Qt}).
For analytical tractability, we denote $\widehat{U}^*$ as the utility obtained by using $\widehat{Q}_i(t)$ for optimization.
Based on (\ref{equ:Q-out-B}), we can prove that the optimality loss due to the outdated knowledge of $\bm{Q}(t)$ is strictly upper bounded as shown in the following theorem.
\begin{theorem}
\label{lemma:O1V2}
Suppose the problem \textbf{P} is feasible, the gap between $\widehat{U}^*$ and $U^\star$ is bounded:
\begin{equation}
U^\star - \widehat{U}^* \leq (B_2+B_3\sum\limits_{\mathclap{{i \in \mathcal{N}} }}\epsilon_i)/V,
\label{equ:Q-out-V}
\end{equation}
where $B_2\!=B_1+\!m \sum\nolimits_{i \in \mathcal{N}}(A_i^{\text{max}})^2$ and $B_3\!=\!2\sum\nolimits_{i \in \mathcal{N}} (c_i^{\text{max}})^2$.
\end{theorem}

\textit{Proof:}~Please refer to  Appendix \ref{section:O1V2}

Theorem \ref{lemma:O1V2} proves that the asymptotic optimality of the proposed ERS-ON algorithm is not compromised with outdated NSI.
Compared with Theorem \ref{lemma:O1V}, the optimality loss caused by outdated NSI is no greater than a constant divided by $V$.
Moreover, the optimality loss is linear to $m$ and increases with the growth of feedback overheads, which coincides with the intuition that a long compulsory feedback interval and a large overhead result in weak system performances.

\begin{table}[t]
\setlength{\abovecaptionskip}{0.cm}
\setlength{\belowcaptionskip}{-3.5cm}
\centering
\caption{System Simulation Parameters}\label{tab:parameters}
\begin{tabular}{|c|c|c|}
\hline
\multicolumn{3}{|c|}{\textbf{Basic Radio Configuration Parameters}}\\
\hline
Spectrum Bandwidth $W$ (MHz) &  \multicolumn{2}{|c|}{0.2~\cite{bibli:3GPP2}} \\
\hline
Pass Loss Exponent $\alpha$ &  \multicolumn{2}{|c|}{2~\cite{bibli:WMEC1}} \\
\hline
$P_0$ / $P_i^{\text{max}}$ / $N_0$ (W) & \multicolumn{2}{|c|}{2/1/$10^{-9}$~\cite{bibli:WMEC1,bibli:3GPP2}}\\
\hline
$e_i^{\text{min}}$ / $e_{i,{\text{max}}}^H$ (mJ) & \multicolumn{2}{|c|}{$0.005/0.16$}\\
\hline
$c_i^{\text{max}}$ / $L$ (Kbits) & \multicolumn{2}{|c|}{100/0.016~\cite{bibli:3GPP1}}\\
\hline
Channel Power Gain & \multicolumn{2}{|c|}{$h_i(t)=10^{-3}d_i^{-\alpha}\widetilde{h}(t)$~\cite{bibli:WMEC1}}\\
\hline
\multicolumn{3}{|c|}{\textbf{Other Simulation Parameters}}\\
\hline
Energy Harvesting Efficiency $\xi_i$ & \multicolumn{2}{|c|}{0.8~\cite{bibli:WMEC1}} \\
\hline
Slot Duration (ms) & \multicolumn{2}{|c|}{100~\cite{bibli:3GPP1}} \\
\hline
Simulation Run-Time (slots) & \multicolumn{2}{|c|}{1000} \\
\hline
\end{tabular}
\vspace*{-11pt}
\end{table}
\section{Simulations}\label{section:sim}
\begin{figure}
\centering{\includegraphics[height=0.35\textwidth]{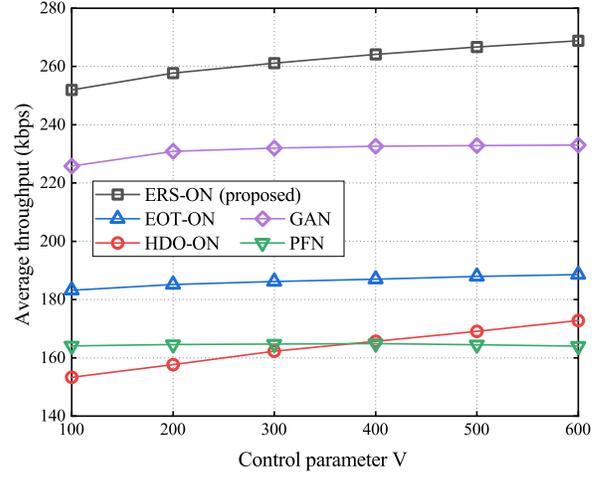}}
\caption{Average system throughput versus the control parameter $V$}
\label{fig:throughput}
\end{figure}
In this section, we verify the analysis and evaluate the performance of the proposed algorithm via MATLAB simulations.
The parameters used in our simulation are taken from the existing synthetic data set~\cite{bibli:WMEC1} and 3GPP specifications~\cite{bibli:3GPP1,bibli:3GPP2}, to capture the features of practical wireless channels.
Detailed information is listed in Table II.
The initial values of $\bm{Q}(t)$, $\bm{S}(t)$ and $\bm{E}(t)$ are set to be zero for the fair comparison with~\cite{bibli:channel}.
In our simulations, we set $T\!=\!100$ms and consider five Type-I devices and five Type-II devices.
Specifically, we set the distance between the AP and Type-I devices as $\bm{d}_1\!=\!(d_i,i \in \mathcal{N}_1)\!=\![3,5,7,9,11]$m.
Without loss of generality, we consider $\bm{d}_2\!=\!(d_i,i \in \mathcal{N}_2)\!=\!\bm{d}_1$.
The channel is modeled after the Rayleigh fading model and is set as $h_i(t)=10^{-3}d_i^{-\alpha}\widetilde{h}(t)$~\cite{bibli:protocol2}, where $\alpha$ denotes the pass-loss exponent, $\widetilde{h}(t)$ is an exponentially distributed random variable with unit mean which represents the short-term fading.
We assume that $\alpha=2$ and the channel reciprocity holds for the uplink and downlink.
The spectrum bandwidth for offloading and the receiver noise power are set as $W\!=\!0.2$MHz and $N_0=10^{-9}$W, respectively.
For any device $i \in \mathcal{N}$, the energy harvesting efficiency $\xi_i \!=\!0.8$, and $A_i^{\text{max}}\!=\!1$Mbps.
At the AP, the transmission power $P_0\!=\!2$W, $r_i^{\text{max}}\!=\!50$kbps, and $m=4$.
For analysis simplicity, we assume $\epsilon_i\!=\!\epsilon\!=\!0.005$s for any $i \in \mathcal{N}$ in the sequel unless otherwise stated.

We introduce four benchmark algorithms to evaluate the performance of the proposed  ERS-ON algorithm, which work as follows:
\begin{enumerate}
  \item Homogeneous device optimization with outdated NSI (HDO-ON): all WDs are seen as Type-I devices and optimized without considering their heterogeneity.
  \item Equal offloading time with outdated NSI (EOT-ON): all WDs are given equal offloading time at each slot.
      The optimization of $\bm{e}(t)$ and $\mu_0(t)$ in this scheme is similar to the ERS-ON algorithm.
  \item Proportional fairness without NSI (PFN):
        the allocation of offloading time portions are based on proportional fairness method with no knowledge of $\bm{Q}(t)$.
        In the scheme, overheads for feedback can be saved.
  \item Greedy algorithm without NSI (GAN): the scheme maximizes the system throughput based on channel conditions without considering fairness.
\end{enumerate}

\begin{figure}
\centering
\subfigure[Fairness performance versus control parameter $V$]{
\label{fig:fairt} 
\includegraphics[width=0.9\linewidth]{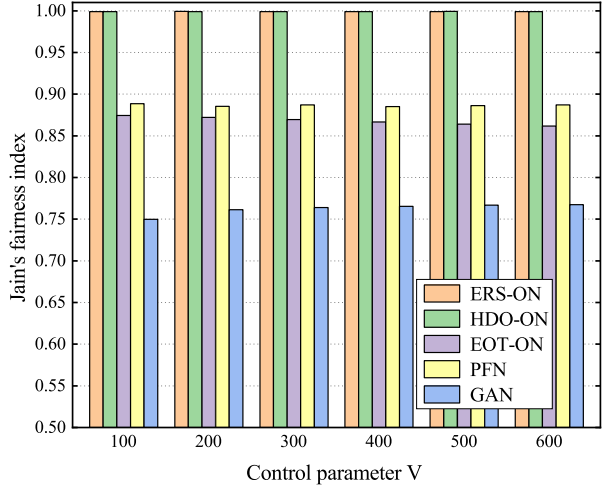}}
\subfigure[Fairness among different device types, where $V\!=\!300$]{
\label{fig:fairi}
\includegraphics[width=0.9\linewidth]{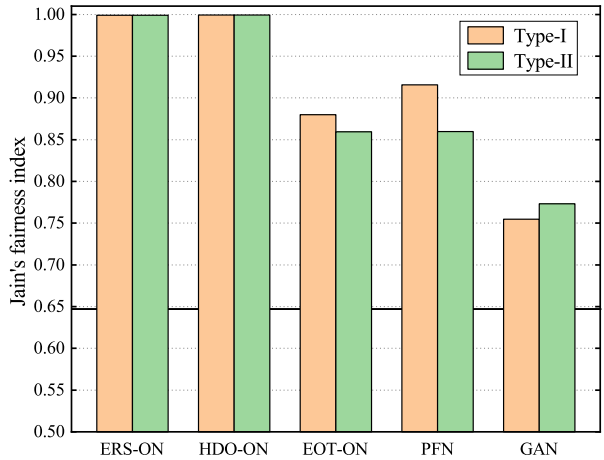}}
\caption{Impact of control parameter $V$ and schemes on fairness}
\label{fig:fairness} 
\end{figure}

Fig. \ref{fig:throughput} illustrates the average throughput of the proposed and benchmark mechanisms as $V$ increases.
Faster growth is seen in the proposed algorithm when $V$ is smaller than 200.
With a further increase of $V$, the growth rate gradually slows down, which coincides with what revealed in Theorem \ref{lemma:O1V2}.
As expected, the proposed algorithm significantly outperforms HDO-ON, since Type-II devices in ERS-ON can save their energy for future use when channel conditions are not good.
This proves the necessity to consider the heterogeneity of WDs in system optimization.
Besides, we can also observe that the ERS-ON algorithm achieves higher throughput than GAN.
This is because GAN wastes much offloading time for WDs that own good channel conditions but with little or even no data backlogs.
The poor performances of GAN and PFN combined suggest that it is desirable to acquire NSI for system optimization even if the feedback would occupy some transmission resource.

\begin{figure}
\centering
\subfigure[Data backlog at the wireless device versus $V$]{
\label{fig:Ql} 
\includegraphics[width=0.9\linewidth]{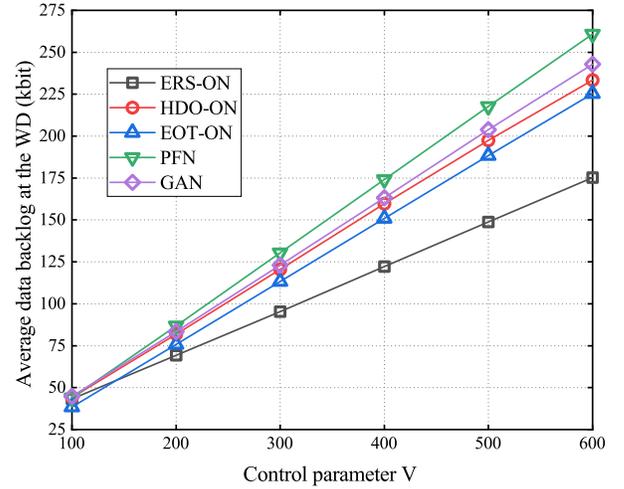}}
\hspace{-0.1in}
\subfigure[Data backlog at the access point versus $V$]{
\label{fig:Sl}
\includegraphics[width=0.9\linewidth]{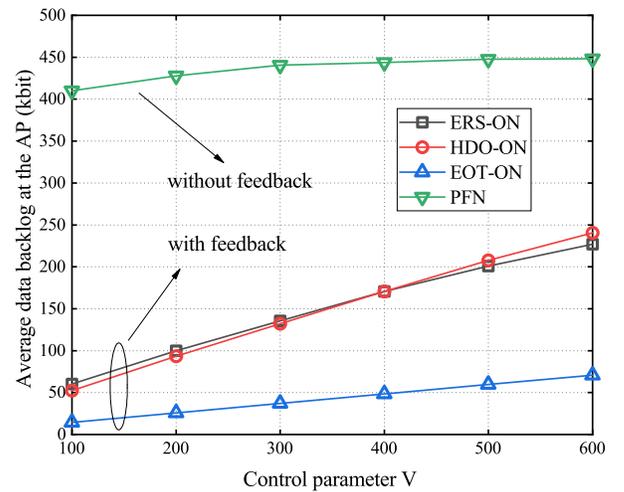}}
\caption{Impact of control parameter $V$ on average data backlog in steady state}
\label{fig:backlog} 
\end{figure}

Fig. \ref{fig:fairness} shows the fairness against the control parameter $V$.
Jain's index is used to measure fairness among WDs \cite{bibli:fairness,bibli:channel}.
We observe from Fig. \ref{fig:fairt} that HDO-ON has almost the same high fairness with ERS-ON algorithm, which is barely affected by the choice of $V$.
However, with less concern on the heterogeneity, HDO-ON achieves the high fairness at the cost of tremendous throughput.
Furthermore, Fig. \ref{fig:fairi} shows the fairness of Type-I and Type-II devices, where $V$ is set as 300.
On one hand, we can see from EOT-ON, PFN, and GAN that heterogeneous battery capacities would lead to an uneven offloading opportunity.
Different from the above schemes, our proposed ERS-ON algorithm strikes a good balance between Type-I and Type-II devices, showing its superiority in handling systems with heterogeneous WDs.
On the other hand, the near to one Jain's index also indicates that ERS-ON can also well tackle the \emph{doubly near-far problem} caused by different $d_i$.

Although higher throughput can be achieved by choosing a larger $V$, it is not always better for the ERS-ON algorithm to do so.
As shown in Fig. \ref{fig:backlog}, the increase of $V$ would also result in linear growth of data backlogs at both the WD and the AP.
By jointly considering Figs. \ref{fig:throughput}, \ref{fig:fairness} and \ref{fig:backlog}, we can observe that the utility-backlog performance follows the [O(1/V),O(V)] tradeoff, which verifies Theorem \ref{lemma:OV} and \ref{lemma:O1V2}.
Besides, the results also provide guidelines for the design of data buffer lengths of WDs in IIoT, since an appropriate $V$ can minimize the cost of embedding lengthy storage spaces given a fixed throughput expectation.
By using ERS-ON algorithm, as depicted in Fig. \ref{fig:Ql}, we can keep the data buffer length at the WD to a small level compared with benchmarks, especially when $V$ is large.
In Fig. \ref{fig:Sl}, the curve for GAN is dropped since its queue length at the AP increases as time goes by and never reaches a steady state.
It can be observed that EOT-ON maintains shorter data backlogs at the AP side.
This is because its volume of offloading data is much less than ERS-ON.
Furthermore, we also observe that scheme with NSI feedback has a shorter data backlog at the AP than one without.

\begin{figure}
\centering{\includegraphics[height=0.36\textwidth]{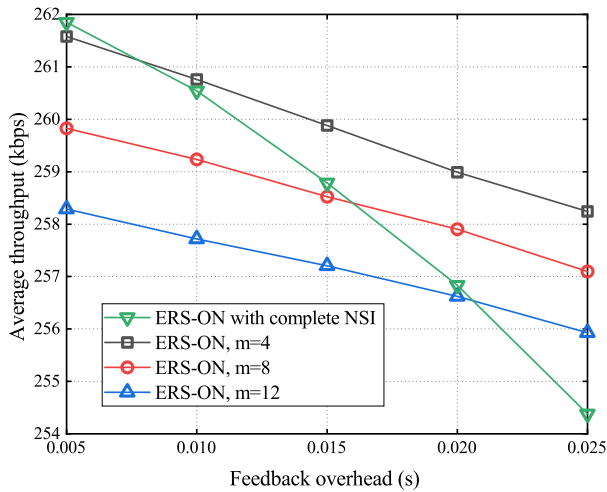}}
\caption{Impact of $\epsilon$ and $m$ on system average throughput}
\label{fig:epsilon}
\end{figure}

For implementation considerations, the impact of $\epsilon$ and $m$ on average throughput is further evaluated.
We set $\epsilon$ to vary from 0.005s to 0.025s, and carry out simulations under different compulsory feedback intervals.
Fig. \ref{fig:epsilon} shows that the average throughput linearly reduces as $\epsilon$ increases.
This is because the NSI feedback process takes up more time originally assigned for offloading when $\epsilon$ becomes larger.
Compared with $m=8$ and $m=12$, ERS-ON with $m=4$ achieves a higher throughput, which corresponds to the conclusion in Theorem \ref{lemma:O1V2}.
Besides, in Fig. \ref{fig:epsilon}, ERS-ON with complete NSI feedback represents that WD has to uplink their NSI in every slot.
Its throughput performance outperforms the others when $\epsilon\!=\!0.005s$.
However, the achievable throughput decreases rapidly as $\epsilon$ grows, and gets worse than ERS-ON with $m\!=\!12$ when $\epsilon\!=\!0.025$s.
This reveals that more NSI contributes to the optimization of resource allocation when the overhead of feedback is small.
Meanwhile, many resources would be wasted by complete NSI feedback when the signaling cost is relatively large.
In this case, it is even better to choose an appropriately large~$m$.

Although the above sections only consider the scenario of a single AP, the proposed algorithm can be readily extended to the scenario of multiple cellular base stations by applying frequency reuse technique~\cite{bibli:interference}.
As the name suggests, the technique utilizes a fraction of the total frequency band in each cell, such that no two neighbor cells use the same frequency.
As a result, inter-cell interference can be reduced, and each cell can run the proposed Algorithm 1 independently with its allocated bandwidth.

\section{Conclusion}
Given the data collection and offloading in WPT-MEC based heterogeneous IIoT systems, we have proposed an energy-aware resource scheduling algorithm for throughput maximization and offloading fairness.
The performance of such an algorithm was analyzed under a typical scenario, where overhead and delay of feedbacks were non-negligible.
Simulations show that the proposed algorithm can achieve asymptotic optimality, and effectively tackle the \emph{doubly near-far problem} as well as the unfairness caused by heterogeneous battery capabilities.
As a result, the use of the algorithm is fully justified by its merits on high throughput, fairness and perpetual operation in implementation.

The integration of the machine-to-machine (M2M) communications to the wireless powered MEC network is promising to tackle the scalability problem in large-scale applications~\cite{bibli:VTM}.
In such a scenario, the proposed algorithm is still valuable.
We can first divide IIoT devices into different clusters according to their proximity and choose the device with the best channel condition in each cell to serve as the cluster head.
The AP only need to schedule those cluster heads for data offloading with the proposed algorithm, while the other devices migrate their data to nearby cluster heads through M2M interactions.
By this way, the computational complexity at the AP can be brought down, and the feedback of NSI can also be reduced.
However, M2M communication brings interference for the data offloaded to the AP.
How to leverage spatial reuse to coordinate the scheduling leaves for future work.

\appendices
\section{}\label{section:Lyapnov}
Note that for any $x \geq 0$, $y \geq 0$, $z \geq 0$, the following inequality
\begin{equation}
([x-y]^+ +z)^2 \leq x^2 + y^2 + z^2 + 2x(z-y).
\label{equ:C2}
\end{equation}
always holds.
Squaring both sides of (\ref{equ:Qt}), summing over $i \in \mathcal{N}$, dividing both sides by $2$ and combining (\ref{equ:C2}), we obtain
\begin{equation}
\begin{aligned}
&~~~~\frac{1}{2}\sum\limits_{i \in \mathcal{N}}[Q_i^2(t+1)-Q_i^2(t)]\\
&\leq \frac{1}{2}\sum\limits_{i \in \mathcal{N}}[c_i^2(t)+a_i^2(t)]+\sum\limits_{i \in \mathcal{N}}Q_i(t)[a_i(t)-c_i(t)]\\
&\leq \frac{1}{2}\sum\limits_{i \in \mathcal{N}}[(c_i^{\text{max}})^2+(A_i^{\text{max}})^2]+\sum\limits_{i \in \mathcal{N}}Q_i(t)[a_i(t)-c_i(t)].
\label{equ:C3}
\end{aligned}
\end{equation}

Similarly, using (\ref{equ:energy-B}) and (\ref{equ:Ft}), we have

\begin{equation}
\begin{aligned}
&~~~~\frac{1}{2}\sum\limits_{i \in \mathcal{N}_2}[(E_i(t+1)-\theta_i)^2\!-\!(E_i(t)-\theta_i)^2] \\
&\leq \frac{1}{2}\sum\limits_{i \in \mathcal{N}_2}[(e_i^H(t))^2\!+\!e_i^2(t)]\!+\!\sum\limits_{i \in \mathcal{N}_2}[E_i(t)\!-\!\theta_i][e_i^H(t)\!-\!e_i(t)]\\
&\leq \frac{1}{2}\sum\limits_{i \in \mathcal{N}_2}[(e_{i,\text{max}}^H)^2+(P_i^{\text{max}}T)^2]\\
&~~+\sum\limits_{i \in \mathcal{N}_2}[E_i(t)\!-\!\theta_i][e_i^H(t)\!-\!e_i(t)],
\label{equ:C4}
\end{aligned}
\end{equation}
and
\begin{equation}
\begin{aligned}
&~~~~\frac{1}{2}\sum\limits_{i \in \mathcal{N}}[S_i^2(t+1)\!-\!S_i^2(t)] \\
&\leq \frac{1}{2}\sum\limits_{i \in \mathcal{N}}[r_i^2(t)\!+\!( \min\{c_i(t),Q_i(t)\})^2]\\
&~~+\sum\limits_{i \in \mathcal{N}}S_i(t)[\min\{c_i(t),Q_i(t)\}\!-\!r_i(t)]\\
&\leq \frac{1}{2}\sum\limits_{i \in \mathcal{N}}[r_i^2(t)\!+\!c_i^2(t)]+\sum\limits_{i \in \mathcal{N}}S_i(t)[c_i(t)-r_i(t)]\\
&\leq \frac{1}{2}\sum\limits_{i \in \mathcal{N}}[(r_i^{\text{max}})^2\!+\!( c_i^{\text{max}})^2]+\sum\limits_{i \in \mathcal{N}}S_i(t)[c_i(t)-r_i(t)]
\label{equ:C5}
\end{aligned}
\end{equation}

Substituting (\ref{equ:C3})-(\ref{equ:C5}) into (\ref{equ:drift}) yield (\ref{equ:bound}).
Taking expectation of the Lyapunov drift conditioned on $\bm{\Theta}(t)$ and subtracting $V\mathbb{E}\{\sum\nolimits_{i \in \mathcal{N}}U_i(a_i(t))\mid \bm{\Theta}(t)\}$ to both sides proves the result.

\section{}\label{section:zero}
By substituting (\ref{equ:energy}) into the objective function of problem \textbf{P2}, the problem can be reformulated into minimizing
\begin{equation}
\begin{aligned}
&\sum\limits_{i \in \mathcal{N}_2}[S_i(t)-Q_i(t)]c_i(t)-\sum\limits_{i \in \mathcal{N}_2}[E_i(t)\!-\!\theta_i]e_i(t)\\
&\!+\!\sum\limits_{\mathclap{ {i \in \mathcal{N}_1} }}[S_i(t)\!-\!Q_i(t)]c_i(t)\!+\!\mu_0(t)\sum\limits_{\mathclap{ {i \in \mathcal{N}_2} }}[E_i(t)\!-\!\theta_i]\xi_i P_0 h_i(t) T.
\label{equ:zero1}
\end{aligned}
\end{equation}
The last term in (\ref{equ:zero1}) a decreasing function of $\mu_0(t)$ since $\sum\nolimits_{i \in \mathcal{N}_2}[E_i(t)\!-\!\theta_i]\xi_i P_0 h_i(t) T$ is a non-positive constant.
For any WD $i \in \mathcal{N}_t \cap \mathcal{N}_1$, we can choose $\mu_i(t)\!=\!0$ to minimize the term $[S_i(t)\!-\!Q_i(t)]c_i(t)$ and re-allocate the originally allocated time portion of WD $i$ for $\mu_0(t)$.
For arbitrary $i \in \mathcal{N}_t \cap \mathcal{N}_2$, as $c_i(t)$ is a non-decreasing function of $e_i(t)$ and $E_i(t)<\theta_i$, the term $[S_i(t)-Q_i(t)]c_i(t)-[E_i(t)\!-\!\theta_i]e_i(t)$ is non-decreasing with $e_i(t)$.
Hence, the minimal value of the first and second terms in (\ref{equ:zero1}) is 0 when $e_i(t)\!=\!0$.
In such condition, $\mu_i(t)$ has no contribution to the value of the objection function.
It is better to make $\mu_i(t)\!=\!0$ and add the value of $\mu_0(t)$ so as to minimize the objective function.

\section{}\label{section:O1V}
Based on \cite{bibli:Lyapunov}, for all $\sigma>0$, there exists a feasible stationary policy $\Pi^\dag$ that chooses $\bm{a}^\dag(t)$, $\bm{e}^\dag(t)$ and $\tilde{\bm{\mu}}^\dag(t)$ such that $\bm{a}^\dag(t)$, $\bm{c}^\dag(t)$ and $\bm{e}^{H\dag}$ satisfy:
\begin{equation}
\begin{aligned}
U^\star\!-\! \mathbb{E}\{\sum\limits_{i \in \mathcal{N}}U_i(a_i^\dag(t))\}&\leq \sigma,\\
\mathbb{E}\{a_i^\dag(t)\!-\!c_i^\dag(t)\} &\leq \sigma,~i \in \mathcal{N},\\
\mathbb{E}\{c_i^\dag(t)\!-\!r_i(t)\} &\leq \sigma,~i \in \mathcal{N}\\
\mathbb{E}\{ e_i^{H\dag}(t)\!-\!e_i^\dag(t)\} &\leq \sigma,~i \in \mathcal{N}_2
\end{aligned}
\label{equ:E1}
\end{equation}

The 0-additive approximation ensures by (\ref{equ:d-p-p}) is as follows
\begin{equation}
\begin{aligned}
\Delta_V(\bm{\Theta}(t))
\leq & B_1 -V \mathbb{E}\left\{\sum\limits_{i \in \mathcal{N}}~U_i\big(a_i^\dag(t)\big)\mid \bm{\Theta}(t) \! \right\}\\
&~+\sum\limits_{i \in \mathcal{N}} Q_i(t)\mathbb{E}\left\{ a_i^\dag(t)-c_i^\dag(t) \mid \bm{\Theta}(t)  \right\}\\
&~+\sum\limits_{i \in \mathcal{N}} S_i(t)\mathbb{E}\left\{ c_i^\dag(t)-r_i(t) \mid \bm{\Theta}(t)  \right\}\\
&~+\sum\limits_{i \in \mathcal{N}_2} \left[E_i(t)\!-\!\theta_i \right]\mathbb{E}\{e_i^{H\dag}(t)\!-\!e_i(t)^\dag\mid \bm{\Theta}(t)\},
\end{aligned}
\label{equ:E2}
\end{equation}

By substituting (\ref{equ:E1}) into (\ref{equ:E2}) and taking $\sigma\rightarrow 0$, we have
\begin{equation}
\Delta(\bm{\Theta}(t))\!-\! V\mathbb{E}\left\{\sum\limits_{i \in \mathcal{N}}~U_i\big(a_i(t)\big)\mid \bm{\Theta}(t) \! \right\} \!\leq\! B_1\!-\!VU^{\star}
\label{equ:E3}
\end{equation}
Using iterated expectations and telescoping sums over $t$, dividing both sides by $Vt$ and rearranging terms yields
\begin{equation}
\frac{1}{t}\sum\limits_{\tau=0}^{t-1}\mathbb{E}\left\{\!\sum\limits_{i \in \mathcal{N}}~U_i\big(a_i(t)\big)\!\right\} \!\geq\! U^{\star}\!-\frac{B_1}{V}\!-\frac{\mathbb{E}\{L(\bm{\Theta}(0))\}}{Vt}
\label{equ:E4}
\end{equation}
Since $\mathbb{E}\{L(\bm{\Theta}(0))\}<\infty$ holds, taking limits on both sides of (\ref{equ:E4}) as $t \to \infty$ proves the result.

\section{}\label{section:OV}
The proof can be obtained by mathematical induction.
We first show that $Q_i(t)$ is deterministic bounded by $V\!+\!A_i^{\text{max}}$ for all $t$.
This clearly holds for $t=0$ as $Q_i(0)=0$.
Suppose it holds for time slot $t$, then we show it also holds for time slot $t+1$.
Consider the case when $Q_i(t) \leq V$.
Since the buffer can increase by at most $A_i^{\text{max}}$ at any slot, we have $Q_i(t+1) \leq V\!+\!A_i^{\text{max}}$.
Then consider the other case when $V < Q_i(t) \leq V+A_i^{\text{max}}$.
In this case, the data admission $a_i(t)$ should choose to be zero according to (\ref{equ:condition1}). Hence, we have
\begin{equation*}
Q_i(t+1)\!=\![Q_i(t)-c_i(t)]^+\! \leq \!Q_i(t)\! \leq \! V+A_i^{\text{max}}.
\end{equation*}
In other words, $Q_i(t+1) \leq V+A_i^{\text{max}}$ holds at $t+1$, which completes the proof of the upper bound of $Q_i(t)$.

Similarly, we proceed to prove $S_i(t)\!\leq\! V\!+\!A_i^{\text{max}}\!+\!c_i^{\text{max}}$ by mathematical induction.
Again, it holds for $t=0$ as $S(t)=0$ at the beginning and we suppose it also holds in time slot $t$.
When $S(t) \leq V+A_i^{\text{max}}$, from (\ref{equ:Ft}), we obtain
\begin{equation}
\begin{aligned}
S_i(t+1) &= [S_i(t) - r_i(t)]^+ + \min\{c_i(t),Q_i(t)\}\\
& \leq S_i(t) + c_i^{\text{max}}\\
& \leq V+A_i^{\text{max}}+c_i^{\text{max}}.
\end{aligned}
\label{equ:Smax1}
\end{equation}
Otherwise, in the case when $S_i(t)\!\geq\! V\!+\!A_i^{\text{max}}$, using the previously proven conclusion (\ref{equ:Qmax}) yields $S_i(t) \geq Q_i(t)$.
This means the corresponding backlog at the AP is larger than the queue length at $i$-th WD, i.e., $i \in \mathcal{N}_t$.
In such a condition, the AP will allocate no offloading time for WD $i$ according to Theorem \ref{lemma:SP}.
Therefore, no data from WD $i$ arrive at the AP at time slot $t$, i.e., $c_i(t)=0$.
We have
\begin{equation}
S_i(t+1)=[S_i(t)-r_i(t)]^+ \leq S_i(t) \leq V+A_i^{\text{max}}+c_i^{\text{max}}.
\label{equ:Smax2}
\end{equation}
From (\ref{equ:Smax1}) and (\ref{equ:Smax2}), we can prove that $S_i(t+1)\!\leq\! V\!+\!A_i^{\text{max}}+c_i^{\text{max}}$ holds at time slot $t\!+\!1$.
This thus concludes the proof.

\section{}\label{section:energy}
If $\theta_i$ is set according to (\ref{equ:theta}) and $E_i(t)\!\leq P_i^{\text{max}}T$ at slot $t$, we obtain
\begin{equation}
E_i(t)-\theta_i \leq - \frac{(V+A_i^{\text{max}})c_i^{\text{max}}}{e_i^{\text{min}}}
\end{equation}
Considering arbitrary WD $i \in \mathcal{N}_2$, we have
\begin{equation}
\begin{aligned}
&[S_i(t)-Q_i(t)]c_i(t)-[E_i(t)-\theta_i]e_i(t)\\
\geq &-Q_i(t)c_i(t)+(V+A_i^{\text{max}})c_i^{\text{max}}\geq 0
\end{aligned}
\label{equ:notime}
\end{equation}
where the second inequality holds because (\ref{equ:Qmax}) is always satisfied.
According to the proof in Theorem \ref{lemma:SP}, WD $i$ will not be allocated time for offloading if (\ref{equ:notime}) holds.
This proves that $E_i(t)>P_i^{\text{max}}T$ is guaranteed under the given $\theta_i$, thus we have $E_i(t)\!+\!e_i^H(t)\!>\!P_i(t)T$.

\section{}\label{section:O1V2}
Before the proof of Theorem \ref{lemma:O1V2}, we first prove the following theorem through mathematical induction.
\begin{theorem}
\label{lemma:difference}
At each time slot, the difference between the backlogs at each WD $i \in \mathcal{N}$ and the AP is strictly bounded, which is given by
\begin{equation}
S_i(t)-Q_i(t) \leq 2c_i^{\text{max}}.
\label{equ:SQd}
\end{equation}
\end{theorem}

\begin{IEEEproof}
The upper bound holds at slot $t\!=\!0$ given that $S_i(t)\!=\!Q_i(t)\!=\!0, \forall i \in \mathcal{N}$. Suppose it holds at time slot $t$. Then, if $S_i(t)\!-\!Q_i(t)< 0$, we have
\begin{equation}
Q_i(t+1)\!=\![Q_i(t)\!-\!c_i(t)]^+ \!+\! a_i(t) \geq Q_i(t)\!-\!c_i(t)\!+\!a_i(t).
\label{equ:Qb}
\end{equation}
Combining (\ref{equ:Smax1}) and (\ref{equ:Qb}), we see that
\begin{equation*}
\begin{aligned}
S_i(t+1)\!-\!Q_i(t+1)&\leq S_i(t)\!-\!Q_i(t) \!+\! 2c_i(t) \!-\!a_i(t)\\
&< 2c_i(t)\!-\!a_i(t) \leq 2c_i^{\text{max}}
\end{aligned}
\end{equation*}
On the other hand, if $0 \leq S_i(t)\!-\!Q_i(t)\leq 2c_i^{\text{max}}$, $c_i(t)\!=\!0$ according to Theorem \ref{lemma:SP}.
Hence, we have
\begin{equation*}
\begin{aligned}
S_i(t+1)\!-\!Q_i(t+1)&= [S_i(t)\!-\!r_i(t)]^+ \!-\!Q_i(t) \!-\!a_i(t)\\
&\leq S(t)\!-\!Q_i(t) \leq 2c_i^{\text{max}}
\end{aligned}
\end{equation*}
As a result, the upper bound also holds at time slot $t\!+\!1$, which concludes the proof.
\end{IEEEproof}

Owing to the feedback mechanism of queue lengths, part of the offloading opportunity will be occupied.
Hence, only $(1- \sum\nolimits_{i \in \mathcal{M}_t}\epsilon_i)T$ can be used for WPT and offloading in each time duration.
This will have a direct influence on $c_i(t)$ and $e_i^H$ as can be seen from (\ref{equ:energy}) and (\ref{equ:capacity1}).
Let $F(t)$ denote the function inside the expectation on the RHS of the drift bound in (\ref{equ:d-p-p}).
Similar to $F(t)$, we are motivated to minimize the expectation of $\widehat F(t)$ under the feedback mechanism, where
\begin{equation}
\begin{aligned}
\widehat F(t) = &-V \sum\limits_{i \in \mathcal{N}}~U_i\big(a_i(t)\big)\\
&~+\sum\limits_{i \in \mathcal{N}} \widehat Q_i(t)[a_i(t)-(1-\! \sum\limits_{i \in \mathcal{M}_t}\epsilon_i)c_i(t)]\\
&~+\sum\limits_{i \in \mathcal{N}} S_i(t)[(1-\! \sum\limits_{i \in \mathcal{M}_t}\epsilon_i)c_i(t)-r_i(t)] \\
&~+\sum\limits_{i \in \mathcal{N}_2} \left[E_i(t)\!-\!\theta_i \right][(1-\! \sum\limits_{i \in \mathcal{M}_t}\epsilon_i)e_i^H(t)\!-\!e_i(t)].
\end{aligned}
\end{equation}

Since the real-time queue length is always larger than the approximate queue backlog, combined with (\ref{equ:Q-out-B}), we have
\begin{equation}
\widehat Q_i(t)\leq Q_i(t) \leq \widehat Q_i(t) + m A_i^{\text{max}}
\label{equ:Fhat}
\end{equation}
Therefore, we can derive that
\begin{equation}
\begin{aligned}
F(t)\!-\! \widehat F(t)\! &=\! \sum\limits_{\mathclap{{i \in \mathcal{N}} }}[Q_i(t)\!-\!\widehat Q_i(t)]a_i(t)\!+\!\sum\limits_{\mathclap{{i \in \mathcal{M}_t} }} \epsilon_i \sum\limits_{\mathclap{{i \in \mathcal{N}} }}S_i(t)c_i(t)\\
&+\sum\limits_{\mathclap{{i \in \mathcal{N}} }}[\widehat Q_i(t)(1- \sum\limits_{i \in \mathcal{M}_t}\epsilon_i)-Q_i(t)]c_i(t)\\
&+\sum\limits_{\mathclap{{i \in \mathcal{M}_t} }}\epsilon_i\sum\limits_{\mathclap{{i \in \mathcal{N}_2} }}[E_i(t)-\theta_i]e_i^H(t)\\
& \leq m\sum\limits_{\mathclap{{i \in \mathcal{N}} }}(A_i^{\text{max}})^2+\sum\limits_{\mathclap{{i \in \mathcal{M}_t} }}\epsilon_i \sum\limits_{\mathclap{{i \in \mathcal{N}} }}[S_i(t)-Q_i(t)]c_i(t)\\
& \leq m\sum\limits_{\mathclap{{i \in \mathcal{N}} }}(A_i^{\text{max}})^2+2\sum\limits_{\mathclap{{i \in \mathcal{N}} }}\epsilon_i \sum\limits_{\mathclap{{i \in \mathcal{N}} }} (c_i^{\text{max}})^2
\end{aligned}
\label{equ:difference}
\end{equation}
where the first inequality holds since $E_i(t)\leq \theta_i$ and (\ref{equ:Fhat}) are satisfied, and the second inequality is derived from Theorem~\ref{lemma:difference}.
Submitting (\ref{equ:difference}) into (\ref{equ:d-p-p}), we have
\begin{equation}
\begin{aligned}
\Delta_V(\bm{\Theta}(t))=&\Delta(\bm{\Theta}(t))\!-\! V\mathbb{E}\left\{\sum\limits_{i \in \mathcal{N}} ~U_i\big(a_i(t)\big)\mid \bm{\Theta}(t) \! \right\}\\
\leq & B_2+B_3\sum\limits_{\mathclap{{i \in \mathcal{N}} }}\epsilon_i +\mathbb{E}\{\widehat F(t)\mid \bm{\Theta}(t) \! \},
\end{aligned}
\label{equ:d-p-p-2}
\end{equation}
where $B_2\!=\!B_1\!+\!m\sum\nolimits_{i \in \mathcal{N}}  (A_i^{\text{max}})^2$ and $B_3\!=\!2\sum\nolimits_{i \in \mathcal{N}} (c_i^{\text{max}})^2$.
We can then prove this theorem by using the similar argument in the proof of theorem~\ref{lemma:O1V}.

\ifCLASSOPTIONcaptionsoff
  \newpage
\fi



%

\bibliographystyle{IEEEtran}
\bibliography{Ref-MEC-TWC}

%




\end{document}